\documentclass[aps,twocolumn,showpacs,preprintnumbers,floats]{revtex4}

\usepackage{graphicx}
\usepackage{dcolumn}
\usepackage{bm}
\usepackage{epsfig}

\begin{document}

\title{The hyperon mean free paths in the relativistic mean field }
\author{
 Q.\  L.\ Wang,
 L.\ Dang,
 X.\ H.\ Zhong\footnote{E-mail: zhongxianhui@mail.nankai.edu.cn}, and
 P. Z. Ning\footnote{E-mail: ningpz@nankai.edu.cn}
       }
\affiliation{
  Department of Physics, Nankai University, Tianjin 300071, China}

\begin{abstract}
The $\Lambda$- and $\Xi^-$-hyperon mean free paths in nuclei are
firstly calculated in the relativistic mean field (RMF) theory.
The real parts of the optical potential are derived from the RMF
approach, while the imaginary parts are obtained from those of
nucleons with the relations: $U^{\mathrm{IY}}_{\mathrm{S}} =
\alpha_{\sigma \mathrm{Y}}\cdot U_{\mathrm{S}}^{\mathrm{IN}}$ and
$U^{\mathrm{IY}}_{\mathrm{V}} = \alpha_{\omega \mathrm{Y}}\cdot
U_{\mathrm{V}}^{\mathrm{IN}}$ . With the assumption, the depth of
the imaginary potential for $\Xi^-$ is $W_{\Xi}\simeq-$ 3.5 MeV,
and for $\Lambda$ is $W_{\Lambda}\simeq-$ 7 MeV at low incident
energy. We find that, the hyperon mean free path decreases with
the increase of the hyperon incident energies, from 200 MeV to 800
MeV; and in the interior of the nuclei, the mean free path is
about $2\sim 3$ fm for $\Lambda$, and about $4\sim 8$ fm for
$\Xi^-$, depending on  the hyperon incident energy.
\end{abstract}



\maketitle

\section{Introduction}

The study of the nucleon/hyperon mean free paths in nuclei/nuclear
matter is an important aspect for us to know about the
nucleon/hyperon in-medium properties, because the nucleon/hyperon
mean free paths relate with the nucleon-nucleon/hyperon-nucleon
interactions in nuclear matter. In nuclear reactions, the
nucleon/hyperon mean free path is a useful concept for summarizing
a large number of experimental data, because the nucleon/hyperon
mean free path in a nucleus can be introduced from the cross
sections of the nucleon/hyperon-nucleus reactions, which can be
obtained from experiment directly.

Since the work of Bethe \cite{bethe} in 1940, there has been
considerable interest in the study of the mean free path of a
proton or neutron in the nuclear medium
\cite{schiffer,collins,negele,Mahau,Li,cheon,HJY,rego,clark,GQLi,Caillon,LWchen}.
However, theoretical calculation on hyperon mean free path in
nuclear medium has never been done due to the scarce experimental
data on hyperon-nucleon/nucleus interactions. Recently, some
experimental progresses on hyperon increased the necessary for the
study of the hyperon mean free path in nuclei. In this paper, we
will firstly attempt to calculate the hyperon mean free paths in
nuclei in the framework of RMF theory.

The RMF theory have been used to study the nucleon mean free paths
\cite{cheon,rego,clark,GQLi}. With RMF method, the nucleon mean
free path was found to be in reasonable agreement with
experimental data. Thus, in present work, we also study the
$\Lambda$ and $\Xi^-$-hyperon mean free paths in nuclei with RMF
method. To calculate the hyperon mean free path in a nucleus, the
essential point is to construct the energy-dependent
hyperon-nucleus optical potentials. In this work, the real scalar
and vector potentials are given by RMF. We attempt to obtain the
hyperon imaginary potentials from those of nucleons
$U_\mathrm{S}^{\mathrm{IN}}$ and $U_\mathrm{V}^{\mathrm{IN}}$ by
the relations: $U^{\mathrm{IY}}_{\mathrm{S}} = \alpha_{\sigma
\mathrm{Y}}\cdot U_{\mathrm{S}}^{\mathrm{IN}}$ and
$U^{\mathrm{IY}}_{\mathrm{V}} = \alpha_{\omega \mathrm{Y}}\cdot
U_{\mathrm{V}}^{\mathrm{IN}}$. In fact, Cooper \emph{et al.} also
suggested to obtain the hyperon imaginary potentials from those of
nucleons by means of multiplying a factor about ten years
ago\cite{cooperj}. The depth of the imaginary potential for
$\Xi^-$-$^{40}$Ca, $W_{\Xi}\approx 3.5$ MeV at
$T_{\mathrm{lab}}$=100 MeV, is compatible with the prediction of
Batty \emph{et al.}\cite{batty1}, $W_{\Xi}\approx 3 $ MeV. We find
that, in the central of the nuclei, the hyperon mean free path is
about $2\sim 3$ fm for $\Lambda$, and about $4\sim 8$ fm for
$\Xi^-$, depending on  the hyperon incident energy. The hyperon
mean free path decreases with the increase of the hyperon incident
energy.

This paper is organized as follows. In the subsequent section we
outline the hyperon-nucleon interactions in RMF. The details of
the complex potentials are described in Sec. \ref{Complex}. Then
the formulas about the hyperon mean free path are deduced in Sec.
\ref{mfp}. The results and discussions are given in Sec.
\ref{result}. Finally a brief conclusion is given in Sec.\
\ref{con}.

\section{The hyperon-nucleon interactions in RMF}\label{HYP}

Within the framework  of RMF theory\cite{ff1,ff2}, the effective
Lagrangian for hyperon and nucleons can be written
as\cite{Tan01,Tan04}
\begin{equation}
{\cal L}={\cal L}_{\mathrm{N}}+{\cal L}_{\mathrm{Y}},\\
\end{equation}
where ${\cal L}_{\mathrm{N}}$ is the standard Lagrangian for
nucleons
\begin{eqnarray}
 {\cal
L}_{\mathrm{N}} &=&
 \bar{\Psi} _{\mathrm{N}}(i\gamma_{\mu}\partial^{\mu}-g_{\omega
\mathrm{N}}\gamma_{\mu}\omega^{\mu} -M_{\mathrm{N}}-g_{\sigma
\mathrm{N}}\sigma\nonumber\\
&&-\frac{1}{2}g_{\rho
\mathrm{N}}\gamma_{\mu}\vec{\tau}^{\mathrm{N}}\cdot\vec{\rho}^{\mu}-e\gamma_{\mu}
\frac{1+\tau_{3}^\mathrm{N}}{2}A_{\mu})\Psi_{\mathrm{N}}\nonumber\\
&&+\frac{1}{2}(\partial_{\mu}\sigma\partial^{\mu}\sigma-m^2_{\sigma})
-\frac{1}{3}g_{2}\sigma^{3}-\frac{1}{4}g_{3}\sigma^{4}\nonumber\\
&&-\frac{1}{4}\Omega_{\mu\nu}\Omega^{\mu\nu}
+\frac{1}{2}m_{\omega}^{2}\omega_{\mu}\omega^{\mu}-\frac{1}{4}\vec{R}_{\mu\nu}\cdot\vec{R}^{\mu\nu}\nonumber\\
&&+\frac{1}{2}m_{\rho}^{2}\vec{\rho}_{\mu}\vec{\rho}^{\mu}
-\frac{1}{4}H_{\mu\nu}\cdot H^{\mu\nu},
 \label{LN}
\end{eqnarray}
with
\begin{eqnarray}
 \Omega_{\mu\nu} &=& \partial_{\nu}\omega_{\mu}-\partial_{\mu}\omega_{\nu},\nonumber\\
\vec{R}_{\mu\nu} &=& \partial_{\nu}\vec{\rho}_{\mu}-\partial_{\mu}\vec{\rho}_{\nu},\nonumber\\
H_{\mu\nu} &=& \partial_{\nu}A_{\mu}-\partial_{\mu}A_{\nu}.
 \label{LN}
\end{eqnarray}
It involves nucleons $(\Psi_{\mathrm{N}})$, scalar $\sigma$ mesons
$ (\sigma)$ , vector $\omega$ mesons $ (\omega_{\mu})$, vector
isovector $\rho$ mesons $ (\vec{\rho}_\mu)$, and the photon $
(A_{\mu})$. The parameter of the nucleonic sector (NL-SH)
 are adopted from Ref. \cite{sharma}, which describes
properties of nuclear matter as well as of finite nuclei
reasonably well. The nucleonic parameter set is presented in Table
\ref{Nparameter}.

\begin{table}
\caption{The parametrization of the nucleonic sector (NL-SH). The
masses are given in (MeV) and the coupling $g_{2}$ in
($fm^{-1}$).} \label{Nparameter} \vspace{0.1cm}
\begin{tabular}{|c|c|c|c|c|c}\hline\hline
\ \ $M_{\mathrm{N}}$\ \ & \ \ \ 939.0  \ \ & \ \  $g_{\sigma \mathrm{N}}$ \ \ \ & \ \ \ 10.444 \ \ \ \  \\
\hline
$m_{\sigma}$ & 526.065 & $g_{\omega \mathrm{N}}$& 12.945 \\
\hline
$m_{\omega}$ & 783.0 & $g_{\rho \mathrm{N}}$ &  4.383 \\
\hline
$m_{\rho}$ & 763.0 &$g_{2}$ &-6.9099\\
\hline
& &  $g_{3}$ & -15.8337  \\
 \hline
 \hline
\end{tabular}
\end{table}

And  ${\cal L}_{\mathrm{Y}}$, the Lagrangian for hyperon yields
\begin{eqnarray}
 {\cal L}_{\mathrm{Y}} &=& \bar{\Psi} _{\mathrm{Y}}(i\gamma_{\mu}\partial^{\mu}-g_{\omega
\mathrm{Y}}\gamma_{\mu}\omega^{\mu} -M_{\mathrm{Y}}-g_{\sigma
Y}\sigma\nonumber\\
&&-\frac{1}{2}g_{\rho
\mathrm{Y}}\gamma_{\mu}\vec{\tau}_{\mathrm{Y}}\cdot\vec{\rho}^{\mu}-e\gamma_{\mu}
\frac{1+\tau_{3,\mathrm{Y}}}{2}A_{\mu})\Psi_{\mathrm{Y}}.\label{LY1}
\end{eqnarray}
where Y  stands for the hyperon $\Lambda$ and $\Xi^-$, and
$\Psi_{\mathrm{Y}}$ is the hyperon field. The Pauli matrices for
nucleons and hyperons are written as $\tau^B_a$ with $\tau^B_{3}$
being the third component.

The (iso)vector coupling constants for hyperon, $g_{\omega
\mathrm{Y}}$ and $g_{\rho \mathrm{Y}}$, are determined from the
constituent quark model (SU(6) symmetry), namely:
\begin{equation}
g_{\omega\Lambda}=2g_{\omega\Xi}=\frac{2}{3}g_{\omega \mathrm{N}},
\end{equation}
\begin{eqnarray}
g_{\rho\Xi}=g_{\rho \mathrm{N}},\ \  g_{\rho\Lambda}&=&0.
\end{eqnarray}
The scalar  coupling constant for hyperon, $g_{\sigma
\mathrm{Y}}$, are determined by fitting to the potential depth of
the corresponding hyperon in normal nuclear density with the
following relation\cite{Tan04}
\begin{eqnarray}
U_\mathrm{Y}&=&g_{\sigma \mathrm{Y}}\sigma^{\mathrm{eq}}+g_{\omega \mathrm{Y}}\omega^{\mathrm{eq}}\nonumber\\
 &=&M_\mathrm{N}\left(\frac{M^*_\mathrm{N}}{M_\mathrm{N}}-1\right)\cdot \frac{g_{\sigma Y}}{g_{\sigma \mathrm{N}}}+
\frac{g^2_{\omega \mathrm{N}}}{m^2_{\omega}}\cdot\frac{g_{\omega
\mathrm{Y}}}{g_{\omega \mathrm{N}}} \rho_0,
\end{eqnarray}
where $U_\mathrm{Y}$ is the hyperon potential depth in normal
nuclear density, $\sigma^{\mathrm{eq}}$ and $\omega^{\mathrm{eq}}$
are the values of $\sigma$ and $\omega_0$ fields at saturation,
and $M^*_\mathrm{N} /M_\mathrm{N}$=0.597 and $\rho_0$=0.146
fm$^{-3}$ for the set NL-SH.

It is well known that the potential well depth of $\Lambda$
hyperon in nuclear matter is about $-30$ MeV, so we use
$U_{\Lambda}=-30$ MeV to obtain the coupling constant $g_{\sigma
\Lambda}$. However, the experimental data on $\Xi^-$ hypernuclei
are very few. Dover and Gal\cite{Dover} analyzed old emulsion data
on $\Xi^-$ hypernuclei and concluded a nuclear potential well
depth of $U_{\Xi}=-21$ to $-24$ MeV. Fukuda et al. \cite{TF}
fitted the very-low-energy part of $\Xi^-$ hypernuclear spectrum
in the $^{12}$C(K$^-$,K$^+$)X reaction and estimated the value of
$U_{\Xi}$ to be  $-16\sim-20$ MeV. Recently, E885 at the AGS
\cite{PK} have indicated a potential depth of $U_{\Xi}=-14$ MeV or
less. Note that these $\Xi^-$ potential depth data are estimated
based on Woods-Saxon potentials. Here, we choose $U_{\Xi}=-16$ MeV
to fix $g_{\sigma \Xi}$. According to the NL-SH parameter set for
nucleons, we easily obtained the coupling constants $g_{\sigma
\Lambda}=6.4694$, $g_{\omega \Lambda}=8.630$, $g_{\sigma
\Xi}=3.2623$ and $g_{\omega \Xi}=4.315$.

\section{complex potentials}\label{Complex}
For the nucleon/hyperon, within the framework of RMF, its Dirac
equation of motion is given by
\begin{equation}
[-i\alpha\cdot\bigtriangledown+U^{\mathrm{B}}_{\mathrm{S}}
)+\beta(M_{\mathrm{B}}-U^\mathrm{B}_{\mathrm{V}} )]\Psi_\mathrm{B}
=E\Psi _\mathrm{B},
\end{equation}
where $U^\mathrm{B}_{\mathrm{S}}$ and $U^\mathrm{B}_{\mathrm{V}}$
are the scalar potential and vector potential for the baryons. B
stands for the baryons N, $\Lambda$, $\Xi^-$, respectively. For
simplicity, we neglect the Coulomb interaction. To describe the
absorptive properties of baryons in nuclei, we set
$U^\mathrm{B}_{\mathrm{S}}$ and $U^\mathrm{B}_{\mathrm{V}}$ are
the complex numbers in present work. Here both of the complex
potentials $U^\mathrm{B}_{\mathrm{S}}$ and
$U^\mathrm{B}_{\mathrm{V}}$ are written as
\begin{eqnarray}
U^\mathrm{B}_{\mathrm{S}}=U_{\mathrm{S}}^{\mathrm{RB}}+iU_{\mathrm{S}}^{\mathrm{IB}}, \label{us}\\
U^\mathrm{B}_{\mathrm{V}}=U_{\mathrm{V}}^{\mathrm{RB}}+iU_{\mathrm{V}}^{\mathrm{IB}}\label{uv}.
\end{eqnarray}
The real parts of the scalar potential and vector potential for
baryons $U_{\mathrm{S}}^{\mathrm{RB}},\
U_{\mathrm{V}}^{\mathrm{RB}}$ are
\begin{eqnarray}
U_{\mathrm{S}}^{\mathrm{RB}} &=&g_{\sigma \mathrm{B}}\sigma_0,\nonumber\\
U_{\mathrm{V}}^{\mathrm{RB}} &=&g_{\omega
\mathrm{B}}\omega_0+g_{\rho
\mathrm{B}}\tau_{3,\mathrm{B}}\rho_{0,3}.
\end{eqnarray}

If we neglected the contributions of isospin to the vector
potential of the hyperons and nucleons, the real part of  the
scalar potential and vector potentials
$U_{\mathrm{S}}^{\mathrm{RY}},\ U_{\mathrm{V}}^{\mathrm{RY}}$ for
hyperons can be related to those of nucleons with
\begin{eqnarray}
U_{\mathrm{S}}^{\mathrm{RY}} &=&\alpha_{\sigma \mathrm{Y}}\cdot U_{\mathrm{S}}^{\mathrm{RN}} \label{sy}\\
U_{\mathrm{V}}^{\mathrm{RY}} &=&\alpha_{\omega \mathrm{Y}}\cdot
U_{\mathrm{V}}^{\mathrm{RN}}\label{vy},
\end{eqnarray}
where $\alpha_{\sigma \mathrm{Y}} = g_{\sigma
\mathrm{Y}}/g_{\sigma \mathrm{N}}$ and $\alpha_{ \omega
\mathrm{Y}} = g_{\omega \mathrm{Y}}/g_{\omega \mathrm{N}}$.



From RMF theory, we can obtain the real parts of the complex
potentials only. To calculate the hyperon mean free path, we have
to know the imaginary part of complex potentials.

Since there is a simple linear relation between the real part of
the hyperon scalar (vector) potential $U_\mathrm{S}^{\mathrm{RY}}$
($U_\mathrm{V}^{\mathrm{RY}}$) and the nucleon scalar (vector)
potential $U_\mathrm{S}^{\mathrm{RN}}$
($U_\mathrm{V}^{\mathrm{RN}}$) (see Eqs.(\ref{sy}) and
(\ref{vy})), the same linear relation should exist in the relation
between the hyperon complex potential $U^\mathrm{Y}_\mathrm{S}$ ($
U^\mathrm{Y}_\mathrm{V}$) and the nucleon complex potential
$U_\mathrm{S}^\mathrm{N}$ ($ U_\mathrm{V}^\mathrm{N}$), which are

\begin{eqnarray}
U^\mathrm{Y}_{\mathrm{S}} &=&\alpha_{\sigma \mathrm{Y}}\cdot U_{\mathrm{S}}^{\mathrm{N}}, \label{sy1}\\
U^\mathrm{Y}_{\mathrm{V}} &=&\alpha_{\omega \mathrm{Y}}\cdot
U_{\mathrm{V}}^{\mathrm{N}}\label{vy1}.
\end{eqnarray}
According to the above suppose, the imaginary  potentials of the
hyperon-nucleus potentials can be derived from those of nucleons
with the following relations:
\begin{eqnarray}
U_\mathrm{S}^{\mathrm{IY}}=\alpha_{\sigma \mathrm{Y}}\cdot U_\mathrm{S}^{\mathrm{IN}} ,\label{ii}\\
U_\mathrm{V}^{\mathrm{IY}}=\alpha_{ \omega \mathrm{Y}}\cdot
U_\mathrm{V}^{\mathrm{IN}}\label{iii}.
\end{eqnarray}
The imaginary parts of the nucleon-nucleus potentials
$U_\mathrm{S}^{\mathrm{IN}}$ and $U_\mathrm{V}^{\mathrm{IN}}$ are
extracted from Ref. \cite{hama} directly, which determined by
fitting a large number of proton elastic scattering data. In fact,
to estimate the imaginary parts of the hyperon-nucleus potentials,
cooper \emph{et al.} firstly assumed that the hyperon imaginary
potentials could be obtained from those of nucleons by means of
multiplying a factor about ten years ago\cite{cooperj}.

In RMF theory, in terms of the scalar potential
$U^\mathrm{B}_\mathrm{S}$ and the vector potential
$U^\mathrm{B}_\mathrm{V}$, the momentum of a baryon propagating in
the nuclei can be determined from
\begin{equation}
E=\sqrt{(M_\mathrm{B}+U^\mathrm{B}_\mathrm{S})^2+k^2}+U^\mathrm{B}_\mathrm{V}\label{energymomentum}
\end{equation}
Replaced the complex potentials $U^\mathrm{B}_\mathrm{S},\
U^\mathrm{B}_\mathrm{V}$ with Eqs.(\ref{us},\ \ref{uv}), the
Eq.(\ref{energymomentum}) can be rewritten as\cite{GQLi}
\begin{equation}
\frac{k^{2}}{2M_\mathrm{B}}+U=E-M_\mathrm{B}+\frac{(E-M_\mathrm{B})^{2}}{2M_\mathrm{B}},
\label{energymomentum2}
\end{equation}
where
\begin{equation}
U=V+iW \label{potential},
\end{equation} with
\begin{equation}
V=U_{\mathrm{S}}^{\mathrm{RB}}+\frac{E}{M_\mathrm{B}}U_{\mathrm{V}}^{\mathrm{RB}}+\frac{1}{2M_\mathrm{B}}
({U^{\mathrm{RB}}_{\mathrm{S}}}^{2}-{U^{\mathrm{RB}}_{\mathrm{V}}}^{2}+{U^{\mathrm{IB}}_{\mathrm{V}}}^{2}
-{U^{\mathrm{IB}}_{\mathrm{S}}}^{2}),
\label{potentialv}
\end{equation}
\begin{eqnarray}
W=U_{\mathrm{S}}^{\mathrm{IB}}+\frac{E}{M_\mathrm{B}}U_{\mathrm{V}}^{\mathrm{IB}}+\frac{1}{M_\mathrm{B}}
(U_{\mathrm{S}}^{\mathrm{RB}}U_{\mathrm{S}}^{\mathrm{IB}}-U_{\mathrm{V}}^{\mathrm{RB}}U_{\mathrm{V}}^{\mathrm{IB}}).
\label{potentialw}
\end{eqnarray}

Note that, $U=V+iW$ can be identified as Schr\"{o}dinger
equivalent potential which is the hyperon optical potential in the
non-relativistic approach, Eq.(\ref{energymomentum}) is identical
to the non-relativistic dispersion relation, except the
relativistic correction, $ (E-M_\mathrm{B})^{2} / 2M_\mathrm{B} $.

It is seen from the Eqs.(\ref{potentialv},\ \ref{potentialw}), the
Schr\"{o}dinger equivalent potentials are energy dependent. At
zero momentum of the hyperon, compared with the usual optical
potential defined in RMF,
$U_{\mathrm{opt}}=U_\mathrm{S}+U_\mathrm{V}$, the Schr\"{o}dinger
equivalent potentials have an additional term, which is the third
term in Eq.(\ref{potentialv}) and (\ref{potentialw}). How much the
third term in Eq.(\ref{potentialv})/ (\ref{potentialw})
contributes to the Schr\"{o}dinger equivalent potentials, which we
will discuss latter.

With the imaginary part of hyperon-nucleus potentials given in
Eqs(\ref{ii},\ \ref{iii}), we obtain the real part of the hyperon
optical potential in Eq. (\ref{potentialv})
\begin{eqnarray}
V&=&U_{\mathrm{S}}^{\mathrm{RY}}+\frac{E}{M_\mathrm{Y}}U_{\mathrm{V}}^{\mathrm{RY}}
+\frac{1}{2M_\mathrm{Y}}({U^{\mathrm{RY}}_{\mathrm{S}}}^{2}-{U^{\mathrm{RY}}_{\mathrm{V}}}^{2}\nonumber\\
&&+\alpha_{ \omega
\mathrm{Y}}^{2}{U^{\mathrm{IN}}_{\mathrm{V}}}^{2}-\alpha_{\sigma
\mathrm{Y}}^{2}{U^{\mathrm{IN}}_{\mathrm{S}}}^{2}),\label{potentialvv}
\end{eqnarray}
and the imaginary  optical potential strength is then obtained in
Eq. (\ref{potentialw}):
\begin{eqnarray}
W&=&\alpha_{\sigma \mathrm{Y}}
U_{\mathrm{S}}^{\mathrm{IN}}+\alpha_{ \omega
Y} \frac{E}{M_\mathrm{Y}}U_{\mathrm{V}}^{\mathrm{IN}}\nonumber\\
&&+\frac{1}{M_\mathrm{Y}}(\alpha_{\sigma \mathrm{Y}}
U_{\mathrm{S}}^{\mathrm{RY}}U_{\mathrm{S}}^{\mathrm{IN}}-\alpha_{
\omega \mathrm{Y}}
U_{\mathrm{V}}^{\mathrm{RY}}U_{\mathrm{V}}^{\mathrm{IN}})\label{potentialww}
\end{eqnarray}

\begin{figure}[ht]
\center
 \epsfig{file=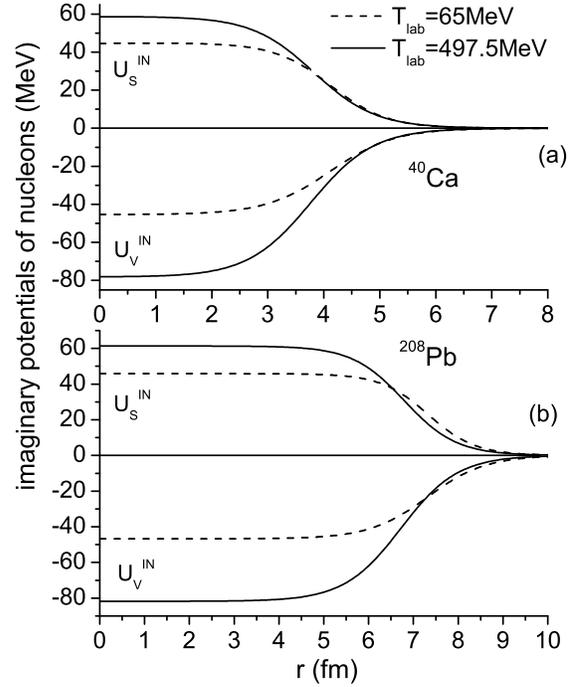,width=8.2cm} \caption{The imaginary scalar potential,
$U_{\mathrm{S}}^{\mathrm{IN}}$ and vector potential,
$U_{\mathrm{V}}^{\mathrm{IN}}$ for N  in $^{40}$Ca and $^{208}$Pb
at the $T_{\mathrm{lab}}$ = 65 and 497.5 MeV, respectively.}
\label{Impp}
\end{figure}

\begin{figure}[ht]
\center
 \epsfig{file=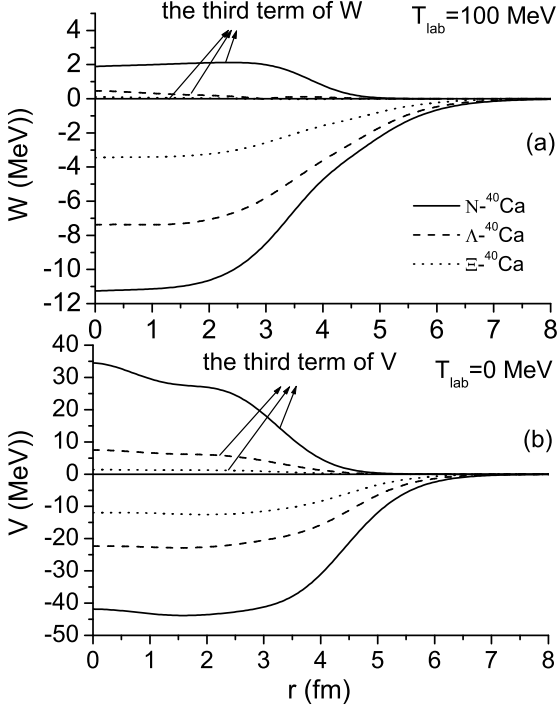,width=8.2cm} \caption{The $\Lambda$-$^{40}$Ca,
$\Xi^-$-$^{40}$Ca and N-$^{40}$Ca potentials are showed. We showed
their real potentials at the $T_{\mathrm{lab}}$ = 0 MeV in Fig.
(b) and their imaginary parts at the $T_{\mathrm{lab}}$ =100 MeV
in Fig. (a). } \label{potentials}
\end{figure}

\section{hyperon Mean free path in RMF}\label{mfp}

Since the potentials $U^\mathrm{Y}_{\mathrm{S}}$ and
$U^\mathrm{Y}_{\mathrm{V}}$ are complex, the hyperon momentum $k$
is also complex and can be expressed as
\begin{equation}
k=k_{\mathrm{R}}+ik_{\mathrm{I}}\label{momentum}.
\end{equation}
The hyperon mean free path, $\lambda$, is related to the imaginary
part of the hyperon momentum by
\begin{equation}
\lambda=\frac{1}{2k_{\mathrm{I}}}. \label{lambda}
\end{equation}
From equations (\ref{energymomentum2}), (\ref{potential}),
(\ref{momentum}) and (\ref{lambda}), it is easy to derive an
analytical expression for $\lambda$,
\begin{eqnarray}
\lambda&=&\frac{1}{2}\bigg\{-M_\mathrm{Y}\left(E-M_\mathrm{Y}-V+\frac{(E-M_\mathrm{Y})^{2}}{2M_\mathrm{Y}}\right)+M_\mathrm{Y}\cdot\nonumber\ \  \\
&&\bigg[\bigg(E-M_\mathrm{Y}-V+\frac{(E-M_\mathrm{Y})^{2}}{2M_\mathrm{Y}}\bigg)^{2}+W^{2}\bigg]^{\frac{1}{2}
}\bigg\}^{-\frac{1}{2}}. \label{lambda1}
\end{eqnarray}
For a  hyperon propagating in a nucleus, its total energy $E$ is
related to the incident energy $T_{\mathrm{lab}}$ by
\begin{eqnarray}
E=\frac{M_\mathrm{Y}^2+m_\mathrm{T}(M_\mathrm{Y}+
T_{\mathrm{lab}})}{[(M_\mathrm{Y}+m_\mathrm{T})^2+2m_\mathrm{T}
T_{\mathrm{lab}}]^{1/2}},
\end{eqnarray}
where $m_\mathrm{T}$ is the mass of nucleus.

\section{results and discussion}\label{result}

In Fig.\ \ref{Impp}, we plot the imaginary parts of the
nucleon-nucleus potentials as functions of radial radii $r$ at
incident energy $T_{\mathrm{lab}}=65$ and 497.5 MeV, respectively.
Obviously, the results of $U_\mathrm{S}^{\mathrm{IN}}$ and
$U_\mathrm{V}^{\mathrm{IN}}$ in Ref. \cite{hama} are repeated.

According to Eqs. (\ref{potentialvv}), (\ref{potentialww}), we can
easily obtain the energy dependent Schr\"{o}dinger equivalent
potentials for the $\Lambda$ and $\Xi^-$  hyperons. As an example,
in Fig. \ref{potentials}, the nucleon, $\Lambda$, $\Xi^-$
Schr\"{o}dinger equivalent potentials in $^{40}$Ca are shown. To
compare with the usual RMF optical potential, we set
$T_{\mathrm{lab}}=0$ and plot the real Schr\"{o}dinger equivalent
potential $V$  in Fig. \ref{potentials} (b). Because the imaginary
nucleon-nucleus potentials are fitted at $T_{\mathrm{lab}}\geq65$
MeV,  we set $T_{\mathrm{lab}}=100$ MeV and plot the imaginary
part $W$ in Fig. (a).

From the figure, we can see that the real parts of the depth of
the Schr\"{o}dinger equivalent potential are $-12$, $-22$ and
$-42$ MeV for $\Xi^-$, $\Lambda$ and nucleon, respectively. The
contributions of the third term in Eqs. (\ref{potentialvv}) to the
real part of the Schr\"{o}dinger equivalent potential are about 7
MeV for $\Lambda$-$^{40}$Ca and even reaches 35 MeV for
N-$^{40}$Ca, however the contributions to $\Xi^-$-$^{40}$Ca is no
more than 1 MeV. As a whole, the depth of the usual RMF optical
potential is deeper than the Schr\"{o}dinger equivalent potential.
For N-$^{40}$Ca and $\Lambda$-$^{40}$Ca, there are large
differences between the usual RMF optical potential and the
Schr\"{o}dinger equivalent potential.

\begin{figure}[ht]
\center
 \epsfig{file=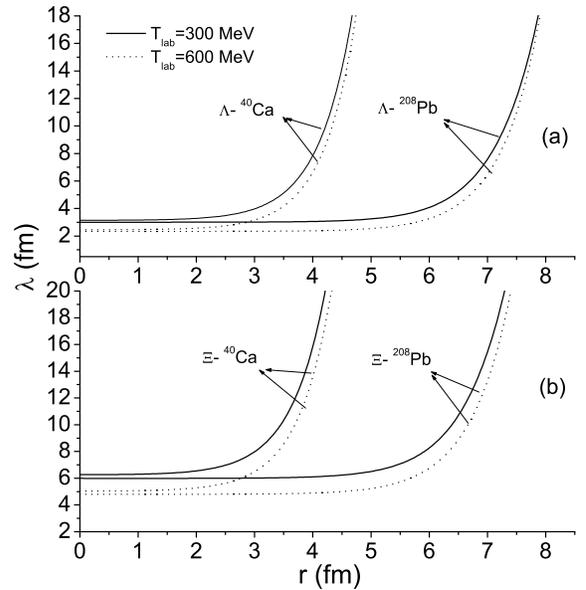,width=8.2cm} \caption{The hyperon mean free
path varied with the radius in $^{40}$Ca and $^{209}$Pb, with the
incident energies, $T_{\mathrm{lab}}$ = 300MeV and 600 MeV. Which
for $\Lambda$ hyperon are showed in Fig. (a) and the $\Xi^-$
hyperon's are shown in Fig. (b).} \label{zong}
\end{figure}

\begin{figure}[ht]
\center
\epsfig{file=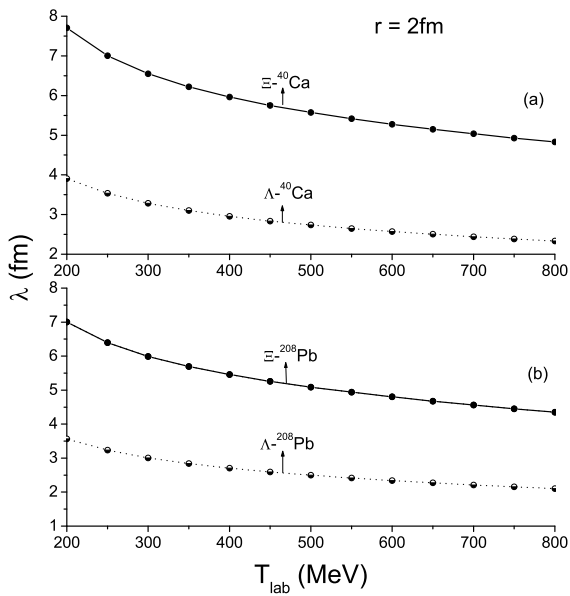,width=8.2cm} \caption{The $\Lambda$,
$\Xi^-$ hyperon mean free path varied with the incident energies,
$T_{\mathrm{lab}}$, from 200 MeV to 800 MeV in $^{40}$Ca and
$^{209}$Pb at $r$ = 2 fm in Fig. (a) and (b) respectively.}
\label{zong1}
\end{figure}

There are few works about the imaginary optical potentials for the
hyperons before. For $\Xi^-$, Dover and Gal suggested an imaginary
part strength of $\sim -1$ MeV\cite{Dover}. While Batty \emph{et
al.} suggested that the depth of imaginary optical potential for
$\Xi^-$ is about 3 MeV, which is calculated with a $t\rho$
potential with Im$b_0$=0.04 fm \cite{batty1}. From the figure, we
find our prediction for depth of the imaginary optical potential
for $\Xi^-$-Ca at low incident energy ($T_{\mathrm{lab}}=$100 MeV)
is $W_{\Xi}\approx -3.5$ MeV, which is compatible with that of
Batty \emph{et al.}. It indicates that our assumption for the
imaginary potentials in Eqs.(\ref{ii}) and (\ref{iii}) is
reasonable. For $\Lambda$, we predicted the depth of the imaginary
potential is $W_{\Lambda}\simeq-$ 7 MeV. The depth of the
imaginary potential for nucleon is $W_{\mathrm{N}}\simeq-$ 11 MeV.
In the imaginary optical potential for $\Xi,\ \Lambda$ and N, the
deepest one is for N, the lowest one is for $\Xi^-$. The
contributions of the last term in Eq. (\ref{potentialww}) to the
imaginary potentials for $\Xi,\ \Lambda$ and N is small, which are
neglectable for $\Xi,\ \Lambda$.

With the determined potentials, we can calculate the hyperon mean
free path in nuclei. We show the hyperon $\Lambda,\ \Xi$ mean free
path in $^{40}$Ca and $^{208}$Pb as function of the radial
distance $r$ in Fig.\ \ref{zong}. The solid, dotted curves
correspond to the hyperon incident energies $T_{\mathrm{lab}}$ =
300 MeV and 600 MeV, respectively. In the center of the nuclei,
the  mean free path  is about $2\sim 3$ fm for $\Lambda$, and
about $5\sim 6$ fm for $\Xi^-$, depending on the the hyperon
incident energy. The mean free path is nearly independent of a
certain nucleus. The hyperon mean free path increases rapidly at
the surface of the nucleus due to the decrease of the
hyperon-nucleus interactions.

To see the effects of the hyperon incident energy  on the mean
free path clearly, we also plot the mean free path in $^{40}$Ca
and $^{208}$Pb (with $r=$2 fm) as a function of incident energy
$T_{\mathrm{lab}}$ in Fig. \ref{zong1}. From the figure, we find
the hyperon mean free path decreases with the increase of the
hyperon incident energy. At $T_{\mathrm{lab}}=$ 200 MeV, the free
mean path in $^{40}$Ca and $^{208}$Pb for $\Lambda$ ($\Xi^-$) is
about 7.8 and 7 (4 and 3.5) fm, respectively, while at
$T_{\mathrm{lab}}=$ 800 MeV, which decreases to about 5 and 4.5
(2.5 and 2) fm.

\section{Conclusion}\label{con}

We have constructed the energy dependent Schr\"{o}dinger
equivalent potentials for hyperon-nucleus, which are functions of
scalar and vector potentials.  The real scalar and vector
potentials for hyperons are obtained from RMF theory directly.
With the assumption,  $ U^\mathrm{Y}_{\mathrm{S}} = \alpha_{\sigma
\mathrm{Y}}U_{\mathrm{S}}^{\mathrm{N}}$ and
$U^\mathrm{Y}_{\mathrm{V}} = \alpha_{\omega
\mathrm{Y}}U_{\mathrm{V}}^{\mathrm{N}}$, we relate the hyperon
imaginary potentials with the nucleon's, which are $
U^{\mathrm{IY}}_{\mathrm{S}} = \alpha_{\sigma
\mathrm{Y}}U_{\mathrm{S}}^{\mathrm{IN}}$ and
$U^{\mathrm{IY}}_{\mathrm{V}} = \alpha_{\omega
\mathrm{Y}}U_{\mathrm{V}}^{\mathrm{IN}}$. The imaginary scalar and
vector potentials, $U_\mathrm{S}^{\mathrm{IN}}$ and
$U_\mathrm{V}^{\mathrm{IN}}$, are obtained from Ref. \cite{hama},
which determined by fitting a large number of proton elastic
scattering data. At low incident energy, the depth of the
imaginary potential $W$ for hyperon-nucleon is on the order of
several MeV. We calculate the depth of the imaginary potential for
$\Xi^-$-$^{40}$Ca, $W_{\Xi}\approx 3.5$ MeV at
$T_{\mathrm{lab}}$=100 MeV, which is compatible with the
prediction of Batty \emph{et al.}\cite{batty1}, $W_{\Xi}\approx 3
$ MeV. The imaginary potential is $W_{\Lambda}\simeq-$ 7 MeV for
$\Lambda$-$^{40}$Ca.

With the determined potentials, we calculated the $\Lambda$,
$\Xi^-$-hyperon mean free path in $^{40}$Ca, $^{208}$Pb with
incident hyperon kinetic energy from 200 MeV to 800 MeV. In the
interior of the nuclei, the hyperon mean free path less depends on
a certain nucleus, and the  mean free path  is about $2\sim 3$ fm
for $\Lambda$, and about $4\sim 8$ fm for $\Xi^-$, depending on
the hyperon incident energy. The hyperon mean free path decreases
with the increase of the hyperon incident energy.

Because there were few works about the hyperon mean free path and
hyperon-nucleus imaginary potential before, the present work  is
only an attempt in this field.  Many theoretical as well as
experimental works are waiting to be done. Besides, the study of
the temperature dependence of the hyperon mean free path are also
needed due to the appearance of hyperons in high-energy heavy-ion
reactions.

\section*{  Acknowledgements }

This work was supported  by the Natural Science Foundation of
China (10275037, 10575054), the Doctoral Programme Foundation of
the China Institution of Higher Education (20010055012) and the
Science and Technology Innovation Foundation of Nankai University,
China. The authors would like to thank Chunyan Song for the useful
discussion, and thank Haixia An for our code.

\end{document}